# Surface Water Formation on the Natural Surface under Super-saturation: from Local Water Balance to Air Pollutant Deposition

Limin Feng[1#], Yang Yu[1], Huan Xie[1], Yujiao Zhu[1], Huiwang Gao[1,2], Xiaohong Yao[1,2*]

[1]Key Lab of Marine Environmental Science and Ecology, Ministry of Education, Ocean University of China, Qingdao, China

[2]Laboratory for Marine Ecology and Environmental Sciences, Qingdao National Laboratory for Marine Science and Technology, Qingdao 266100, China

*corresponding authors: Xiaohong Yao, email: xhyao@ouc.edu.cn

#OCID: https://orcid.org/0000-0002-0962-0404

# Abstract

Heterogeneous nucleation and subsequent growth of surface water occur on the natural substrate when the water vapor concentration reached the point of super-saturation. This study focuses on the parameterization of super-saturation on the canopy-air interface by field observations monitoring surface water formation (SWF) such as dew and frost in the evergreen shrub at an urban cite during autumn and winter in 2015-2017. Here we show that both the interfacial and vertical temperature differences ranged from 1 to 3 K and were necessary but not sufficient for super-saturated condensation on the natural surface. Excessive supplies of moisture must exist, continuously contribute to the growth of the condensed water embryos, originate from both the local and the external sources such as evapotranspiration and atmospheric advection driven by the reduced air pressure, cause SWF not only on the ground soil but also on the vegetation canopy at 1-2 m height. The super-saturation ratio is mainly determined by the coefficient of thermophoresis deposition, which approaches to 1. SWF on the natural surface is not only an indicator but also a weak cleaner of air pollution. The downward thermophoresis deposition of fine particle and droplets favors SWF and the scavenging of air pollutants. The removal efficiency of the deposition flux during SWF event for [$SO_4^{2-}$+$NO_3^-$] is estimated by ~0.3 mmol (per [$Ca^{2+}$] meq)/$m^2$(per leaf area).

**Key words**: surface water formation, water balance, super-saturation, particle deposition.

# 1. Introduction

Atmospheric water accounts for only 0.001% of the Earth's available water, but it is one of the most unstable components, changes in different forms such as water vapor, cloud/fog, rain/snow, dew/frost and aqueous aerosols, plays a ubiquitous role in the Earth's climate system[1,2]. Phase transitions of atmospheric water arise from the contradiction between the actual amount of water vapor in the atmosphere and the capacity of atmosphere to hold water vapor, the latter known as *saturation water vapor pressure (f)*. Water vapor diffuses on a substrate, forms a cluster of nucleuses (heterogeneous nucleation). When the water vapor pressure is continuously greater than the *f*, called *super-saturation*, the nucleated water embryo will keep growing, leading to the formation of surface water on the substrate. This necessitates that the temperature of the substrate be lower than that of the surrounding atmosphere or, more precisely, that the *f* at the substrate be smaller than the *f* at ambient air[3]. However, even this kind of basic physical relationship remains uncertain because some parameters are hard to measure, such as surface free energy, contact angle, surface tension of the boundary phase (an intermediary between the gas and liquid/solid phases). According to the *Clausius-Clapeyron* equation, to determine the *f* value, one must assume that the latent heat is the function of air temperature, which is not the case under the natural condition.

Aerosols can take in the water vapor without reaching super-saturation due to hygroscopicity[4]. The transmit of atmospheric water to and from the particles in the near-surface air is negligible, thus cannot affect the ambient water vapor content. However, the natural surface, such as the vegetation canopy, is not an ideal substrate for water vapor condensing on that because of hydrophobicity, unless the ambient water vapor concentration reaches super-saturation. Therefore, the large amount of water vapor transforms to liquid or solid surface water on the natural surface such as dew/frost, which could affect the local water balance. In order to balance this part of the water vapor loss, there must be a supplemental source. The sources of moisture supply include the atmospheric advection and local evapotranspiration[5,6]. Locally observed high nucleation rate on the canopy-air interface must be the result of small-scale horizontal advection of moisture[6]. Thus, the dew/frost is also called "atmospheric wells", which is one of the important freshwater sources in arid regions. Determining the relative contribution from different moisture sources to the super-saturated condensation is always a challenging target[7]. In addition to water vapor uptake, the wet natural surface also acts as the sink of gases and aerosols[8]. Gas diffusion to the surface water and particle thermophoresis play a key role on the downward deposition flux, the latter caused by vertical temperature gradient at 0-2 m height from the ground. Radiative cooling and the heat loss of the ground generate this kind of surface-based temperature inversion during a clear night.

Global climate change and human activities affect local water balance, and have an impact on the air pollutant deposition. Though intensive researches have advanced our knowledge, the interrelationships involved in local water balance and air pollutant deposition are frequently confusing. (a) In the context of global warming, generally, terrestrial atmosphere becomes drier while marine atmosphere becomes wetter. Without going to detail, a cartoon figure of global water balance under climate change can be seen in Fig. 1. (b) Inorganic salts in the aerosols such as sulfate and nitrate released by intensified natural and anthropogenic sources promote water uptake both in the sub-saturated and super-saturated regimes[9]. Therefore, the increase of the concentration of fine particles in the atmosphere may change the water balance between the ground and atmosphere. (c) Reduced surface winter northerlies and enhanced thermal stability of the lower atmosphere occur more frequently under climate change[10], which is conducive to a relatively warm and humid weather condition. These abnormal weather patterns are favorable for extreme air pollution and surface water formation (SWF) on the natural surface.

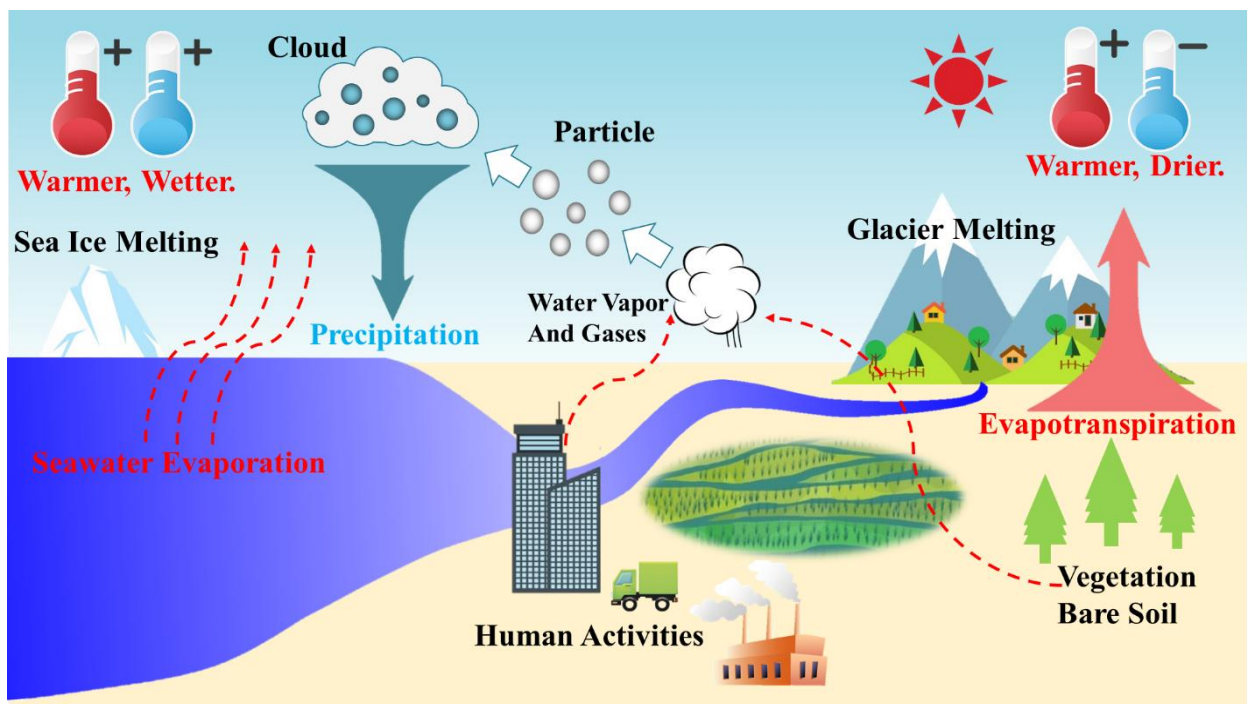

Fig. 1 Macroscopic View: global water balance under climate change.

Considering the potential important role of SWF on local water balance and air quality, we still need clearer information about the thermodynamic and kinetic mechanisms governing super-saturated condensation. Urban forest landscape was an important site for dew deposit[11]. This additional input of water can be of great relevance to the local water balance and air quality[12]. From November to December of 2015 and from December 2016 to January 2017, the *Arctic Oscillation Index* is positive, indicate that cold Arctic air was locked in the polar region, and the relatively warm and humid air prevailed over the North China Plain. During these periods, we observed that SWF events and air pollution events were springing up in an urban site. This article focuses on the following two questions: (a) On the canopy-air interface, how to use the observation of dew and frost to express the super-saturation and so to quantify the impact of SWF on local water balance? (b) How does SWF contribute to the deposition and removal of air pollutants? Fig. 2 cartoons a concept image to outline this article's scope.

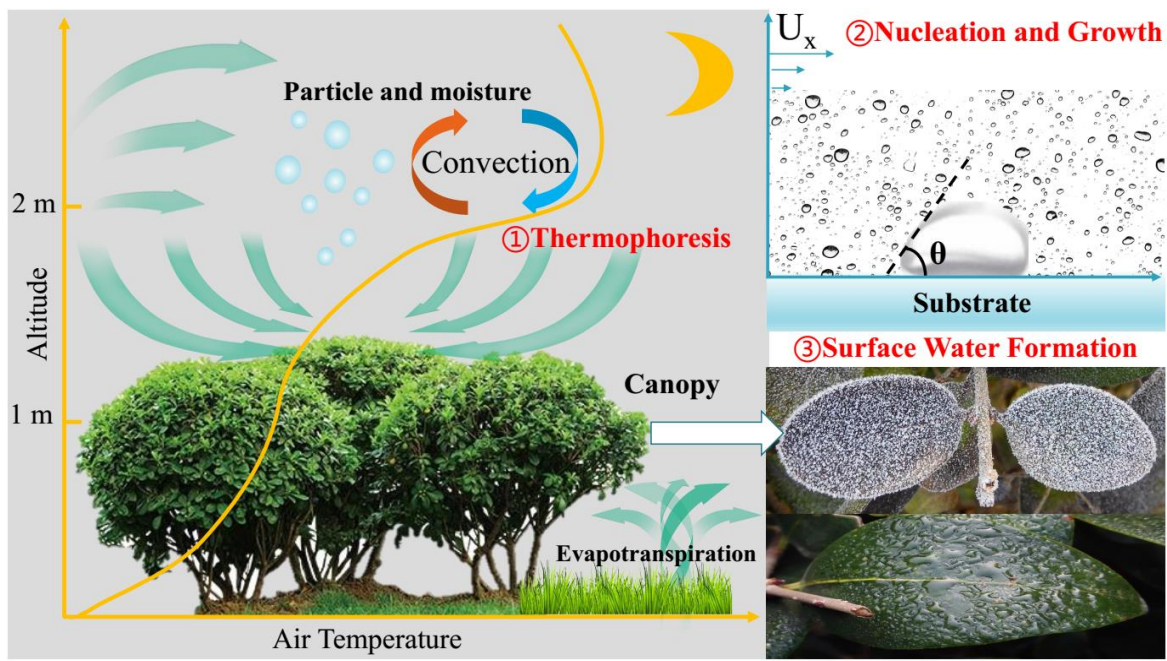

Fig. 2 Microcosmic View: local water balance. ①Thermophoresis deposition of aerosols and moisture onto the topsoil and vegetation canopy; ②Nucleation and growth on substrate, $0° < \theta < 180°$; ③ Surface water formation (SWF), dew and frost.

## 2. Method

**Event records**. The field observation took place in an evergreen shrub at an urban site in Qingdao (36.161 °N, 120.496 °E). We use a camera and a cold light lamp to shooting the formation of dew and frost on the shrub canopy during the night and morning. The shrub canopy is about 1 m height above the ground. We catch 37 dew/frost events from 2015-10-20 to 2017-01-18. To make a concise description, for example, we named an event as (D, F) which means dew appearing on the ground soil and frost appearing on the shrub canopy (D=dew, F=frost, N=none).

**Meteorological parameters.** Air temperature (AT), relative humidity (RH) and wind speed were measured by automatic weather stations near the shrub, with three temperature sensors placed at 0.1 m, 1.0 m, 2.0 m height above the ground. 3-D wind speed was measured by Eddy Covariance (EC 150, Campbell Scientific Ltd., USA). Surface temperature (ST) of the shrub and ground was measured by two infrared temperature probes (IR816A, Bokles Corporation) connected to a data logger. The probes were placed at 1.5 m height above the ground, while the horizontal position of the probes varied in different nights to shoot on different leaves and topsoil. The precision of the infrared temperature probe is ±2%. The measurements of surface temperature of soil and leaf may be influenced by the object distance between canopy and probe.

**Online measurement of water vapor and aerosols.** We set up several instruments to measure the concentrations of water vapor, particles and gases. A greenhouse-gas analyzer (GGA, Model 911-0011, LGR Corporation) was placed near the shrub to measure the water vapor concentration (ppm). The gas inlet is at 2 m height above the ground. An optical particle sizer spectrometer (OPS, Model 3330, TSI

Corporation) was used for scanning the particle size spectrum in 0.3-10 μm and to get the particle volume concentration ($10^6$ μm/m$^3$). An ambient ion monitor coupled with an ion chromatograph (AIM−IC, URG-9000, Thermal Scientific) was set up to measure the concentrations of water-soluble inorganic ions (μg/m$^3$) in PM$_{2.5}$. The instruments used to collect aerosols were indoor with inlet at 5 m height above the ground. However, our aerosol instruments sometimes were not available. The governmental monitoring station (120.459 °N, 36.085 °E) is about 6 km away from our sampling site. Thus we get daily published PM$_{2.5}$ mass concentration (μg/m$^3$) data from this station (See website: http://www.qepb.gov.cn/slairday.aspx).

**Off-line chemical analysis of inorganic ions.** When dew and frost appear on the shrub leaves, we used Milli-Q Water (18.2 Ω, Res. J Scientific Instruments Cooperation) to wash out them to the test tubes, and then the solution passed through 0.45 μm filter to remove insoluble impurities. Each test tube contains solution (about 40 mL) that washed out from 20 leaves. Then the solution was diluted with Milli-Q water to 50 mL volumetric flask. These solution samples were firstly stored in a frozen refrigerator (-20℃) and then extracted for the subsequent chemical analysis. The mass of water-soluble inorganic ions ($K^+$, $Cl^-$, $SO_4^{2-}$, $NO_3^-$, $Na^+$, $Ca^{2+}$, $Mg^{2+}$, $NH_4^+$) in the samples were detected by Ion Chromatograph (IC-1100, ICS-1100, Dionex, USA). Wash-out operations were carried out in the early morning and afternoon to get solution samples both from wet leaves and dry leaves. Each leaf was washed out just for one-time and then cut off to ensure no repetition. This kind of wash-out operation may be affected by subjective selection of shrub leaves varied with different sizes, thus normalization is a must. Normalization includes three steps: (a) Calculate the 20 leaves' total area by outline drawing on the standard gridding paper and convert the ion mass (μg) to concentration with canopy area (μg/m$^2$);

(b) Convert the ion concentration to the mole equivalent concentration (meq, μmol/m$^2$); (c) Divide the meq of $SO_4^{2-}$, $NO_3^-$, $NH_4^+$ by the meq of $Ca^{2+}$. For a further explanation, $Ca^{2+}$ mainly derives from crustal sources such as dust with high concentrations comparative to nitrate and sulfate. $Ca^{2+}$ is non-volatile, thus the $Ca^{2+}$ mass should be nearly constant on the equal quantity of leaves at the same time. $SO_4^{2-}$, $NO_3^-$, $Ca^{2+}$, $NH_4^+$ account for ~ 80% mass of the total mass of inorganic ions in the wash-out solution of dew/frost. The primary source of $K^+$, $Cl^-$ is biomass burning. $Na^+$ mainly comes from the sea salt aerosols. Therefore, we regard $K^+$, $Cl^-$, $Na^+$ as natural background. $Mg^{2+}$ has the same characteristics as $Ca^{2+}$, but part of the $Mg^{2+}$ may originate from marine sources in coastal areas. And $Mg^{2+}$ concentration is low, the detection error of the instrument may be larger, so this article doesn't use $Mg^{2+}$ to normalize the other ions's concentration. We get ions' concentration (mg/kg) in the topsoil (0-1 cm depth) by wash-out operation that is the same as wash-out operation of leaf. Details of online and offline methods above can be found in former article[13-15].

**Statistical analysis of the data.** Although the aerosols' measurements are fixed-point, the measurement of the surface temperature of and the wash-out operation were randomly choosing leaves. Therefore, normal distribution test is required to ensure the homogeneity of the samples. In this article, we adopted Quantile-Quantile (Q-Q) plot to test homogeneity. If the scatters' distribution is clearly divided into two or more groups, it indicates that the data violates the homogeneity. On top of that, we adopted Z-score standardization, that is, $x^* = \dfrac{x-\mu}{\sigma}$, where μ= mean value, σ= standard deviation. Data that beyond the 3 times of σ is treated as an outlier and is excluded.

# 3. Result and Discussion

## 3.1 Parameterization of super-saturation on the canopy-air interface

### 3.1.1 Moisture Supply

A routine fitting between water vapor concentration ($C_{H_2O}$) and air temperature (AT) at 2 m height can be seen in the Fig. 3(a). All scatters are nocturnal mean values averaged between 18:00-6:00 (Beijing Time, the same below). During the nights without any SWF (dew or frost) on canopy (either soil or vegetation), i.e., (N, N) events, the relationship between $C_{H_2O}$ and AT is: $C_{H_2O} = 3515e^{0.09AT}$; While during the nights with SWF on canopy, i.e., (D, D) and (F, F) events, the relationship between $C_{H_2O}$ and AT is: $C_{H_2O} = 4614e^{0.13AT}$. Comparing with (N, N) events, SWF required water vapor content increased by 30% in the dew/frost events. In the nights with lower water contents, even air temperature is close to 0℃, none surface water was formed on the canopy. One might expect that the diurnal temperature range (DTR) will perform a role in SWF. But considering examples in the two events of 2016-11-14 and 2016-11-15 when DTR > 10 ℃, dew appeared on the canopy in the former night, while not the case for the latter one. Thus absolute value of AT is not the governing factor of super-saturated condensation on the canopy, nor the DTR. Actually, the vegetation canopy could lower the daytime surface temperature, reduce the heat loss of the ground surface at night by latent heat releasing, and reduce the DTR due to its effect on microclimate.

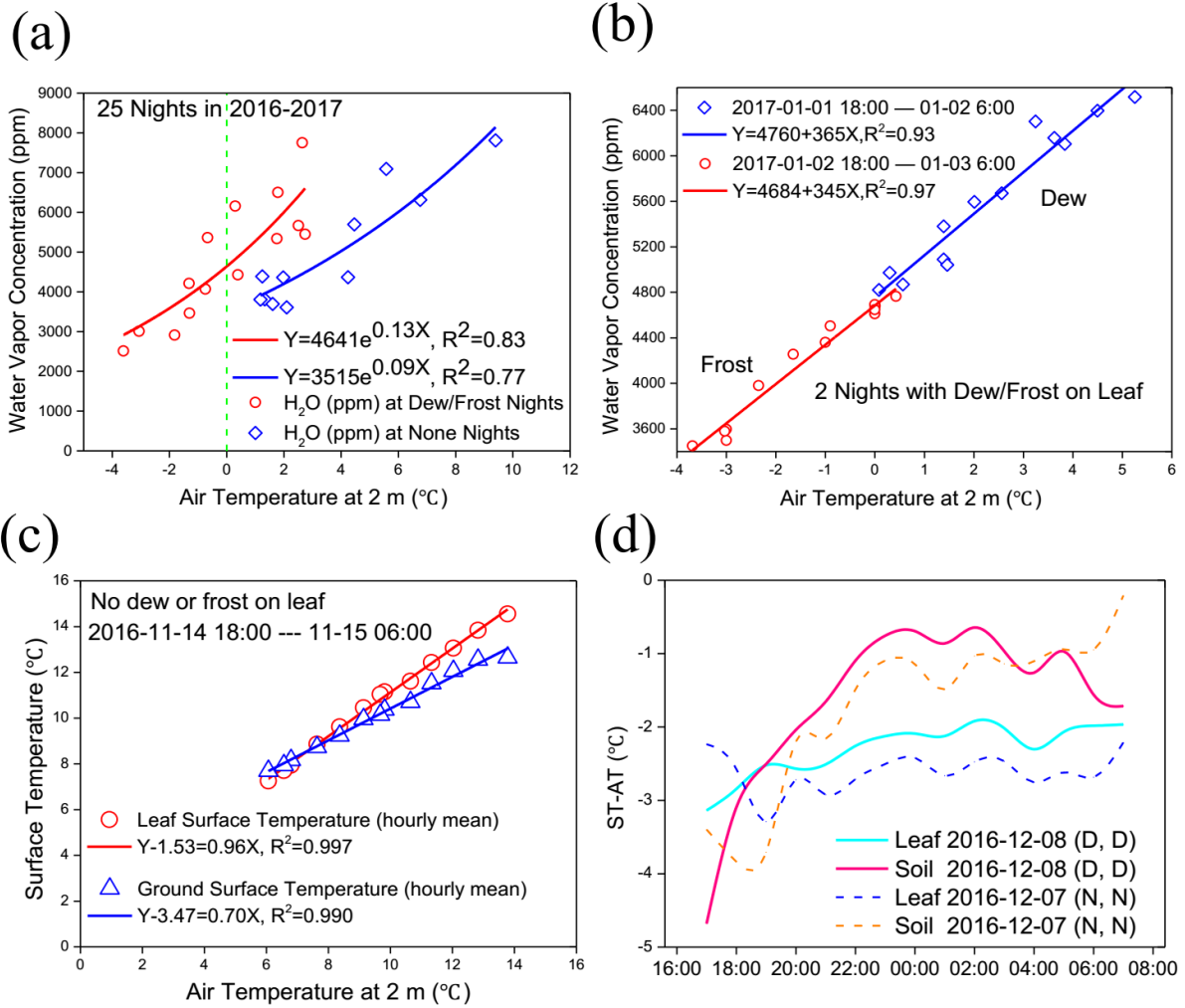

Fig. 3 Regression between meteorological parameters: (a) Regression between Air Temperature (AT) and Water Vapor Concentration, nighttime mean (18:00-6:00) value; (b) Regression between Air Temperature and Water Vapor Concentration, hourly mean value at night; (c) Regression between Air Temperature and Surface Temperature (ST), including leaf and topsoil; (d) Interfacial temperature difference (ST-AT) in (D, D) and (N, N) events.

During a (D, D) or (F, F) events, ambient water vapor concentration may change in three different ways. (i)Assuming an extreme condition that the condensation area is in a closed system without any local or external moisture supplies. SWF depletes a significant portion of atmospheric moisture, thus $C_{H_2O}$ will decline suddenly when SWF begins while irrelevant to nocturnal AT decreasing. (ii) There are significant moisture supplies, which replenish the moisture depletion caused by condensed nucleus growth. In this case, the ambient air reaches a balance between moisture depletion and supply, thus $C_{H_2O}$ will decrease in accordance with AT decreasing while not influenced by SWF. (iii) The local moisture supplies are robust, which is excessive compared to the moisture depletion by SWF. $C_{H_2O}$ will increase during the night and is not influenced by AT nor by SWF. See a typical example in the Fig. 3(b) to verify these ways. During the 2 nights of 2017-01-01 to 2017-01-03，surface water appeared on the shrub canopy and ground soil. We get linear relationships of AT versus $C_{H_2O}$ with the slope ~350, the intercept ~4700 ppm which is the same as the exponential function in Fig. 3(a). Thus in these nights the depletion and replenishment of atmospheric moisture were comparable, and the local water vapor content reached a balance with AT decreasing.

We regard the linear function $C_{H_2O}^{\circ}$ = 4600+350AT as a threshold to judge whether there are significant moisture supplies, and compare $C_{H_2O}$ with $C_{H_2O}^{\circ}$ during different types of events. See the Fig. 4(a). During (F, N) events, i.e., frost appearing on soil but not on shrub canopy, we show that $C_{H_2O}$ < $C_{H_2O}^{\circ}$ in the nighttime, and indicate that the local moisture supply is weak, which mainly come from the soil. During the (F, F) and (D, D) event, i.e., frost/dew appearing both on soil and shrub canopy, we show that $C_{H_2O} \approx C_{H_2O}^{\circ}$ in the nighttime. Thus local water supply almost completely replenishes the water vapor deficit caused by SWF. An interesting episode occurred during 2017-01-03 to 2017-01-04, when $C_{H_2O} > C_{H_2O}^{\circ}$, indicate that the there was an excessive local moisture supply dominating the local water vapor content, which must come from external sources. In the Fig. 4(b), we get similar results with Fig. 4 (a). Surface water appearing on the shrub canopy is indispensable to the excessive moisture supplies that may come from external sources, while appearing on the ground just need water vapor that comes from the soil moisture. In the (N, N) events, the water vapor concentration was comparable with the (F, N) events and less than 4000 ppm, which indicate that the absolute value of water vapor concentration is vital important to SWF on vegetation canopy. It also suggests that in addition to the moisture supply, there may be other factors controlling SWF on the ground, such as wind speed. Strong winds obstruct radiative cooling and condensation[16].

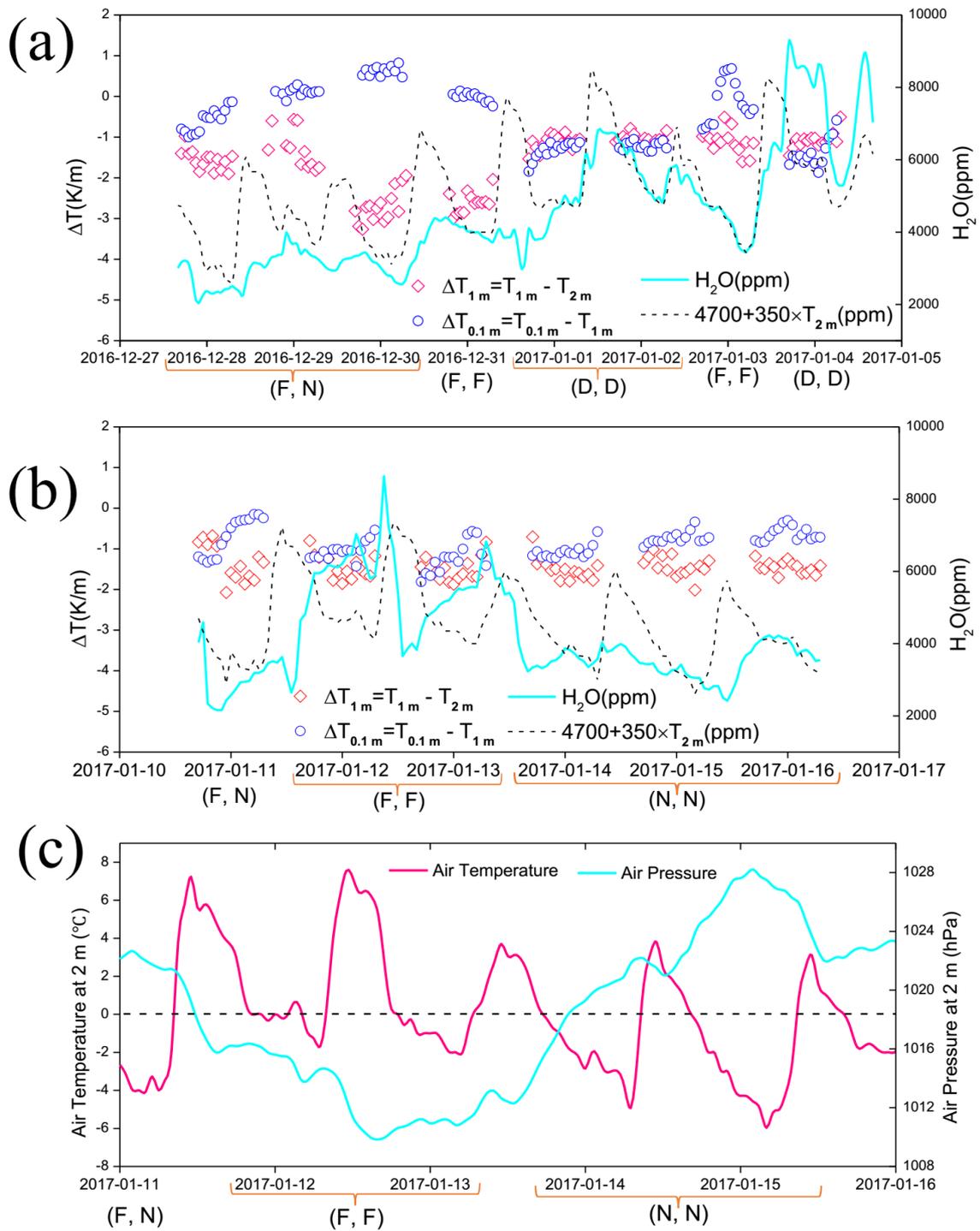

Fig. 4 Temporal variation of water vapor concentration and vertical air temperature gradient: (a) Period-A; (b) Period-B; (c) temporal variation of air pressure and air temperature at 2 m height during frost events.

The different ratio of water vapor concentration ($C_{H_2O}/C_{H_2O}^\circ$) during (D/F, D/F) and (D/F, N) events indicates that the contributions of different moisture sources. During the daytime of 2017-01-01 to 2017-01-03, Fig. 5 shows that the fitting curve between AT versus $C_{H_2O}$ during nighttime is above the fitting curve between AT versus $C_{H_2O}$ during daytime. The increase of water vapor concentration during the day is slower than the decrease of water vapor concentration during the night with AT change. This contradiction is due to 60% interception water by natural surface was re-evaporate to the atmosphere, with the left part of water pass through the soil or vegetation, saved as soil moisture[5]. When the water vapor concentration is relatively low ($C_{H_2O} < C_{H_2O}^\circ$) that couldn't cause SWF on the vegetation canopy, the soil moisture still contributes to the condensation on the soil-air interface, which needs a period of water interception or absorption in advance, either from fog, rain, and dew/frost. When the water vapor concentration is relatively high ($C_{H_2O} > C_{H_2O}^\circ$), the excessive moisture supplies cause SWF on the vegetation canopy and ground, and replenish the deficit of soil moisture. Therefore, SWF is a buffer between atmosphere and ground and facilitate the local water balance between the atmosphere and ground.

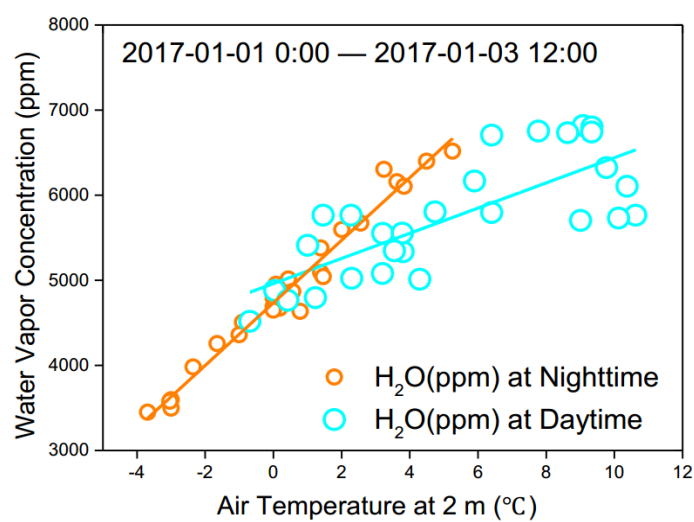

Fig. 5 Regression between Air Temperature and Water Vapor Concentration, hourly mean value at daytime and nighttime.

We notice that when surface water appeared on the shrub canopy, so did the ground soil, i.e., (D, D) and (F, F) events. However, when surface water appeared on the ground soil, chances are that there was no surface water on shrub canopy, i.e., (F, N) and (D, N) events. Water vapor evaporates from the deeper soil pore and condense on topsoil's fallen leaf or grass before reach shrub canopy. Some earlier studies concerning about the relationships of dew and crop production have showed that the evaporated moisture from soil could not reach a certain height ($< 1$ m)[17]. Thus even moisture supply come from the external advection is weak, water vapor still reaches super-saturation on soil-air interface, while not the case for shrub-air interface which needs significant moisture supply. Besides, soil exhibit a hygroscopic effect, whereby they absorb water before the onset of dew[18]. But natural canopies with fatty contents displayed a smaller or negligible hygroscopic effect. In other words, super-saturation is easier to reach in the soil-air interface than in the shrub-air interface. Thus the RH at 0.1 m height is typically higher than that at 2 m height (see in the Fig. 6).

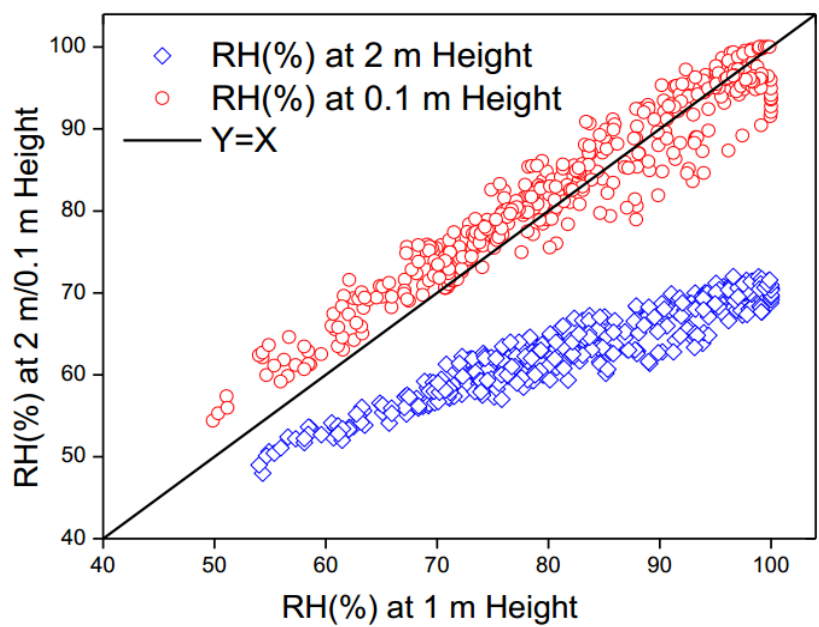

Fig. 6 RH at different height.

The pathways that excessive moisture supplies reach to the condensation area is mysterious. Water vapor condensation in cloud is generally found to accompany with the drop of air pressure in 3-4 km height[19]. Back to the Fig. 4(b), the water vapor concentration was continuously increasing during SWF, like (F, F) events in the nights of 2017-01-12 and 2017-01-13. However, the air pressure was decreasing during former night while increasing during the latter night. See in the Fig. 4(c). Note that the air pressure was decreased prior to the (F, F) events in 2017-01-12. Decreased air pressure could cause a small-scale advection, which contain sufficient moisture from external area to replenish even increase local water vapor content. An additional focus can be seen in the Fig. 7. During the night of 2017-01-03 to 2017-01-04, with a continuously decreasing AT, air pressure was increasing before frost formation (about 00:00) while decrease after condensation begun. A record video of surface water formation during this episode can be seen in my YouTube website: https://youtu.be/fElhK8wb2gw. Air pressure at the surface depend on the amount of gas molecules in the column. Therefore, the air pressure was increasing when the water vapor concentration was low in (N, N) events. Nonetheless, there is doubt as to whether SWF leads to reduced air pressure, or more precisely, does the decreased air pressure contribute to or just a result of super-saturated condensation? Condensation may influence air pressure through the mass removal of atmospheric water accompany with the latent heat release, which may in turn enhance the moisture supply and give a positive feedback to condensation. The weak advection is favorable for condensation, since strong flow will disturb the nucleation and subsequent growth. The wind speed of sampling site at 2 m height was lower than 1 m/s in dew/frost events (See in the Fig. 7(a)). Advections are much smaller in the vertical direction than in the horizontal plane. One possible scene is that the super-saturated condensation induces a horizontal air pressure gradient. Since the air pressure is related to the wet air density, a horizontal water vapor gradient was formed and drive the external moisture supply to the condensation area.

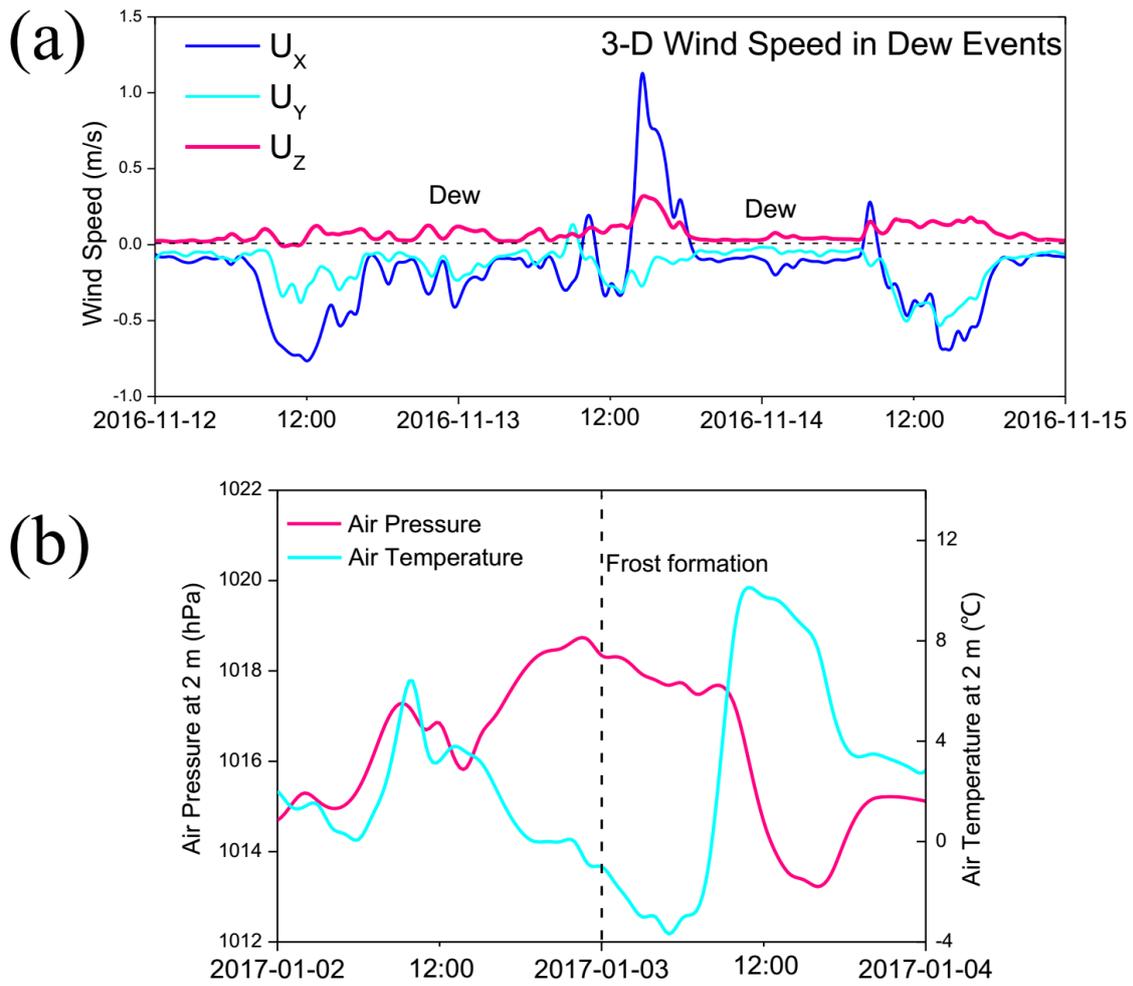

Fig. 7 (a) 3-D wind speed in dew events; (b) temporal variation of air pressure and air temperature at 2 m height during a frost event on 2017-01-03.

**3.1.2 Interfacial and vertical temperature differences**

When there is a clear dry atmosphere above, there is stronger longwave radiation cooling of the surface, which gives a lower minimum surface temperature at night[20]. Radiative cooling lowers the surface temperature of substrate with respect to air temperature by a value less than 3 K[16]. Temperatures of leaves exposed to clear skies after sunset were 1–3 K lower than those of the adjacent air[21], but leaves without surface water were 1–2 K colder than those with that[22]. Considering that the surface temperature of the objects is controlled by the ambient air temperature, we use the air temperature at the 1.0 m to calibrate the surface temperature of the shrub leaf, and use the air temperature at the 0.1 m to calibrate the surface temperature of the topsoil. An example is showed in Fig. 3(c). We use the AT to calibrate the skin surface temperature (ST) of soil and canopy. We get a linear relationship: $ST - b = aAT$. The coefficient $b$ is the measuring error of infrared probes, and is deducted in the final calculation. For soil, $0.5 < a < 0.7$; For shrub canopy, $0.8 < a < 1.0$. The $a$ values indicate that the surface temperature of shrub canopy is close to ambient air temperature ($T_{1m}$), while the surface temperature of ground soil is lower than ambient air temperature ($T_{0.1m}$), explained by the release of the ground heat to the ambient air. During the night of 2016-11-14 18:00 to 2016-11-15 06:00, there is no surface water on the canopy, even though the ST of both soil and shrub canopy was lower than AT.

Fig. 3(d) shows that during a (D, D) event, interfacial temperature difference (|ST-AT|) was 1-2 K lower than that in a successive (N, N) event. From the prospect of surface heat balance in the night, we get: G↑=LW↑+LE↓ (LW=longwave radiation, G=ground heating, LE=latent heat, ↑=net outgoing, ↓=net downward). In a night with clear sky, the ground loses heat by outgoing longwave radiation, but gets heat from latent heat released by condensation. Free convection did not contribute significantly to heat transfer in the nocturnal boundary layer[22]. Thus super-saturated condensation is not favorable for a significant interfacial temperature difference between canopy and air. Our observation confirms that interfacial temperature difference in the canopy-air interface was small (1-2 K) and is necessary but not sufficient for water vapor content reach super-saturation on canopy surface.

The radiative cooling causes a lower ST and AT on ground than that on overlying air, forming a temperature gradient above the ground. This kind of thermal stratification formed at ground level where dew/frost is formed[23]. Former studies have confirmed the existence of surface-based temperature inversion at 0-2 height above the ground, especially in the night of autumn and winter typically with a depth of 1-2 m[24]. We define vertical temperature difference (K/m) as: $\Delta T_{1m} = T_{1m} - T_{2m}$ and $\Delta T_{0.1m} = T_{0.1m} - T_{1m}$. Recall the Fig. 4(a)(b), which shows the temporal change of $\Delta T_{1m}$ and $\Delta T_{0.1m}$ during the night with surface water formation. We do not show the daytime data for a concise view. During all nights, $\Delta T_{1m} < -1$ K, in other words, $T_{1m} < T_{2m}$. While $\Delta T_{0.1m}$ varied with different types of SWF events. In (D, D) nights, $\Delta T_{1m} \approx \Delta T_{0.1m} < 0$, i.e., $T_{0.1m} < T_{1m} < T_{2m}$. In (F, N) and (F, F) events, $\Delta T_{1m} < \Delta T_{0.1m}$, even $\Delta T_{0.1m} > 0$, i.e., $T_{0.1m} > T_{1m}$. Latent heat of sublimation is bigger than latent heat of vaporization, thus the latent heat flux released by the ground soil offset more radiative cooling effect during the frost formation on canopy surface. In the (N, N) events with clear sky, typical temperature inversion were formed with $T_{0.1m} < T_{1m} < T_{2m}$. Similar as the interface temperature difference, vertical temperature difference is also a necessary but not sufficient condition for super-saturated condensation.

When the temperature inversion occurred near the surface, upper water vapor contains higher kinetic energy in intense heat diffusion. Therefore, vapors collide with particle, push the fine particle and droplets move across the air temperature gradient. This kind of particle movements is called thermophoresis deposition, with particle deposition velocity is proportional to the temperature gradient[25]: $V = \lambda \frac{-\Delta T}{AT}$. The thermophoresis coefficient $\lambda$ depends on the properties of both the gas and the particle, such as kinematic viscosity of the gas. Temperature gradient also induces a water vapor gradient, cause water vapor moving from the upper air to the ground. It should be note that deposition velocity of vapors include turbulence in addition to the gradient force. Thus this linear relationship may have some uncertainties when it comes to the application of the vapor deposition velocity. However, considering that the shrub leaf is covered by hydrophobic fatty substance, water vapor itself is not enough to cause the initial nucleation on the canopy by molecule diffusion, thus a significant amount of droplets must accumulate on the leaf beforehand. For fine particles in 0.1-2.5 μm, gravitational settling was of smaller magnitude than the turbulent deposition, and the effect of Brownian diffusion is negligible normally, and the movement of a particle is affected by convection within the gas flow and thermophoresis only [18]. Since the convection is generally weak during SWF, the vertical movements of droplets and particle are governed by the thermophoresis.

### 3.1.3 Brief Summary: Parameterization

The key point of super-saturation is that there is a difference of capacity between air and natural surface to contain water vapor when excessive moisture continuously supplies to maintain this premium above the $f$ and cause the nucleation rate is more than zero. The nucleation rate is determined by many factors, such as contact angle and surface free energy. But these parameters are hard to measure exactly. See the following function[3]: $\frac{dn}{dt} = \exp\left( k \frac{\gamma^3}{\Delta E^2} \times \Phi(\theta) \right)$. When $\theta \approx \frac{\pi}{2}, \Phi(\theta) \approx 1$, and $k \frac{\gamma^3}{\Delta E^2}$ is related to the interfacial temperature difference. In this article, recalling the fitting functions in the Fig. 3 and Fig. 4, we use an experimental function to express the super-saturation ratio: $f^* = \Delta S \times e^{\theta(AT\text{-}ST)}, f^* > 1 \Leftrightarrow \frac{dn}{dt} > 0$. The $\Delta S$ means enhanced water vapor flux, and it is the result of overlying moisture supply and downward water vapor movement, thus we get: $\Delta S = \Delta C_{H_2O} \times V$. Now we get the following relationships:

$$f^* = \Delta S \times \frac{dn}{dt}$$

$$f^* = \Delta S \times \exp[\theta(AT - ST)]$$

$$\theta = 0.13$$

$$\Delta S = \Delta C_{H_2O} \times V$$

$$V = Turbulance \approx \lambda \frac{-\Delta T}{AT}$$

$$\Delta C_{H_2O} = \frac{C_{H_2O}}{C^\circ_{H_2O}}$$

$$C^\circ_{H_2O} = 4700 + 350 \times (AT - 273.15)$$

Assuming that the kinematic viscosity $\lambda$ is constant, we can calculate the super-saturation ratio. We show the six examples of SWF events in the Table. 1 with measured parameters and calculated super-saturation. Video records are also presented as YouTube web links. We get $f^* \sim 0.01\lambda$, and uncertainties come from failing to consider the turbulence of water vapor. It is believed that in natural conditions, $f^* < 0.01$. Thus we take the $\lambda$ value as 1. In the former studies, $\lambda$ value ranged from 0.2-1.2 under the laboratory condition[25]. Thus further studies need to verify our assumption.

Table.1 The measured parameters and calculated *Super-Saturation* ratio.

| Condensation Begin | Surface Water | $ST - AT$ | $\frac{C_{H_2O}}{C^\circ_{H_2O}}$ | $\Delta T_{1m}$ | $f^*$ |
|---|---|---|---|---|---|
| 1-2016-12-07 23:00 | Droplet | -2.1 | 1.04 | -2.3 | $0.011\lambda$ |
| 2-2017-01-02 00:00 | Droplet | -1.1 | 1.05 | -1.1 | $0.005\lambda$ |
| 3-2017-01-03 00:00 | Ice | -1.1 | 1.04 | -1.0 | $0.004\lambda$ |
| 4-2017-01-03 21:00 | Droplet | -1.3 | 1.46 | -1.0 | $0.007\lambda$ |
| 5-2017-01-12 04:00 | Ice | -1.6 | 1.39 | -1.5 | $0.009\lambda$ |
| 6-2017-01-13 03:00 | Ice | -1.2 | 1.39 | -1.4 | $0.008\lambda$ |

* **YouTube links**:

1-2016-12-08 https://youtu.be/UYQNfet8DP4

2-2017-01-02 https://youtu.be/d9-HgXp61ow

3-2017-01-03 https://youtu.be/_gzgpuH_PX8

4-2017-01-04 https://youtu.be/fElhK8wb2gw

5-2017-01-12 https://youtu.be/n4AIWgP2NnM

6-2017-01-13 https://youtu.be/6mnIdd0xGgI

## 3.2 Air Pollutant Deposition on canopy during SWF

### 3.2.1 SWF as an indicator

Since SWF occurred on a thin nocturnal boundary layer above canopy with a weak advection, dew events usually accompany with weather conditions conducive to air pollution, such as surface-based temperature inversion and a humid warm air[23], related to the clear sky radiation as well as enormous heat transfer from ground to the surface air. Here we sort the $PM_{2.5}$ concentrations from the smallest to the largest during dew/frost events, as showed in Fig. 8. When there is no SWF on shrub canopy, $PM_{2.5}$ concentrations were taken as a negative number. In the first quadrant, $PM_{2.5}$ concentrations in (F, F) events in significantly larger than that in (D, D) and (F, N) event. However, there are still three (D, D) events with a large $PM_{2.5}$ concentrations (>150 μg/m$^3$). We standardized these two parameters of $PM_{2.5}$ concentration and air temperature. These three (D, D) events occurred with a low air temperature. Both AT and $PM_{2.5}$ are one time SD lower than mean value in (F, N) events. However, in (F, F) events, low AT accompany with a high $PM_{2.5}$ concentrations which is 1 times SD larger than mean value. Thus, this discrepancy reconfirms that the controlling factors of (F, N) and (F, F) events are different, explained by different portion of moisture supplies from soil and advection. Excessive moisture supplies cause a humid weather condition, which may have a significant effect on PM$_{2.5}$ concentration. We can see that SWF on canopy under low air temperature (-4-4 ℃) occurred with air pollution.

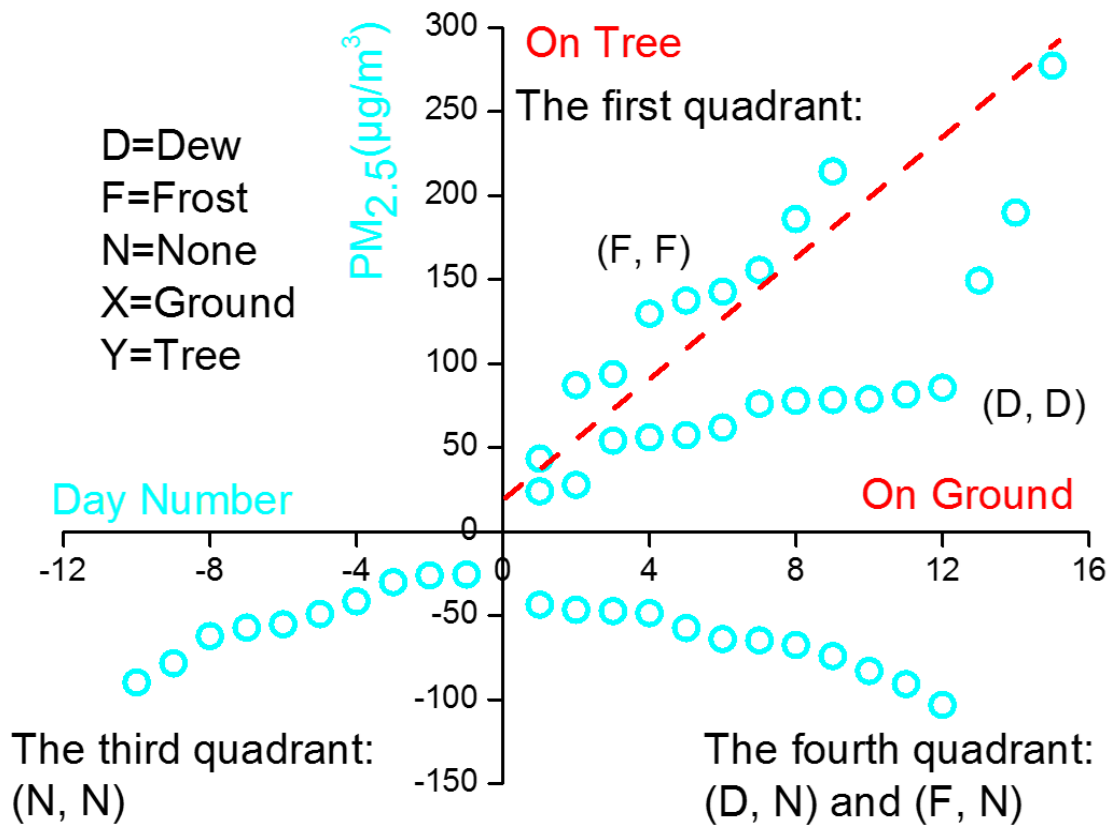

Fig. 8 Correlations between dew/frost events and air pollution events. PM$_{2.5}$ concentrations in 5 categories of dew/frost events in 2015-2017, with PM$_{2.5}$ concentrations from lowest to highest in each category.

Another supporting evidence can be seen in the Fig. 9(a). PM $_{0.3\text{-}1\ \mu m}$ and PM $_{1\text{-}2.5\ \mu m}$ reached a high pitch when dew and frost appearing on the shrub canopy (Peak 1-3). Still, Fig. 9(b) shows the temporal trend of mass concentrations of NH$_4^+$, NO$_3^-$, SO$_4^{2+}$ in PM$_{2.5}$. NH$_4^+$, NO$_3^-$, SO$_4^{2+}$ concentrations reached the peak in the morning and begin decreasing after sunrise, explained by the disappear of radiation-induced temperature inversion layer. The high particle concentration is not a causal factor for SWF, because the high particle concentrations also occurred in (N, N) events. However, chances are that the thermophoresis deposition during SWF could scavenge part of the air pollutants accumulated in the nocturnal boundary layer. See a clearer example in the Fig. 9(c). During a frost event (4:00-7:00) with high particle concentration (PM$_{2.5}$>200 μg/m$^3$), the increase of NH$_4^+$, NO$_3^-$, SO$_4^{2+}$ concentration in the PM$_{2.5}$ was slowing down. After sunrise, nitrate and sulfate concentrations in PM$_{2.5}$ continue to increase until the temperature inversion and air pollutant input were receded in afternoon. A lower air temperature and a higher water content are conducive to gas to particle process such as nitrate formation. Temperature inversion leads to the elevated fine particle concentration and a relatively weak scavenging effect during SWF caused by thermophoresis deposition. Thus our next goal is quantifying the amount of deposition flux of fine particle and gas by collecting dew/frost and analyzing their inorganic composition. One might argue that gravitational settlement of course particles must play a role in clearing air pollutant, but nitrate and sulfate are generally concentrate in fine mode.

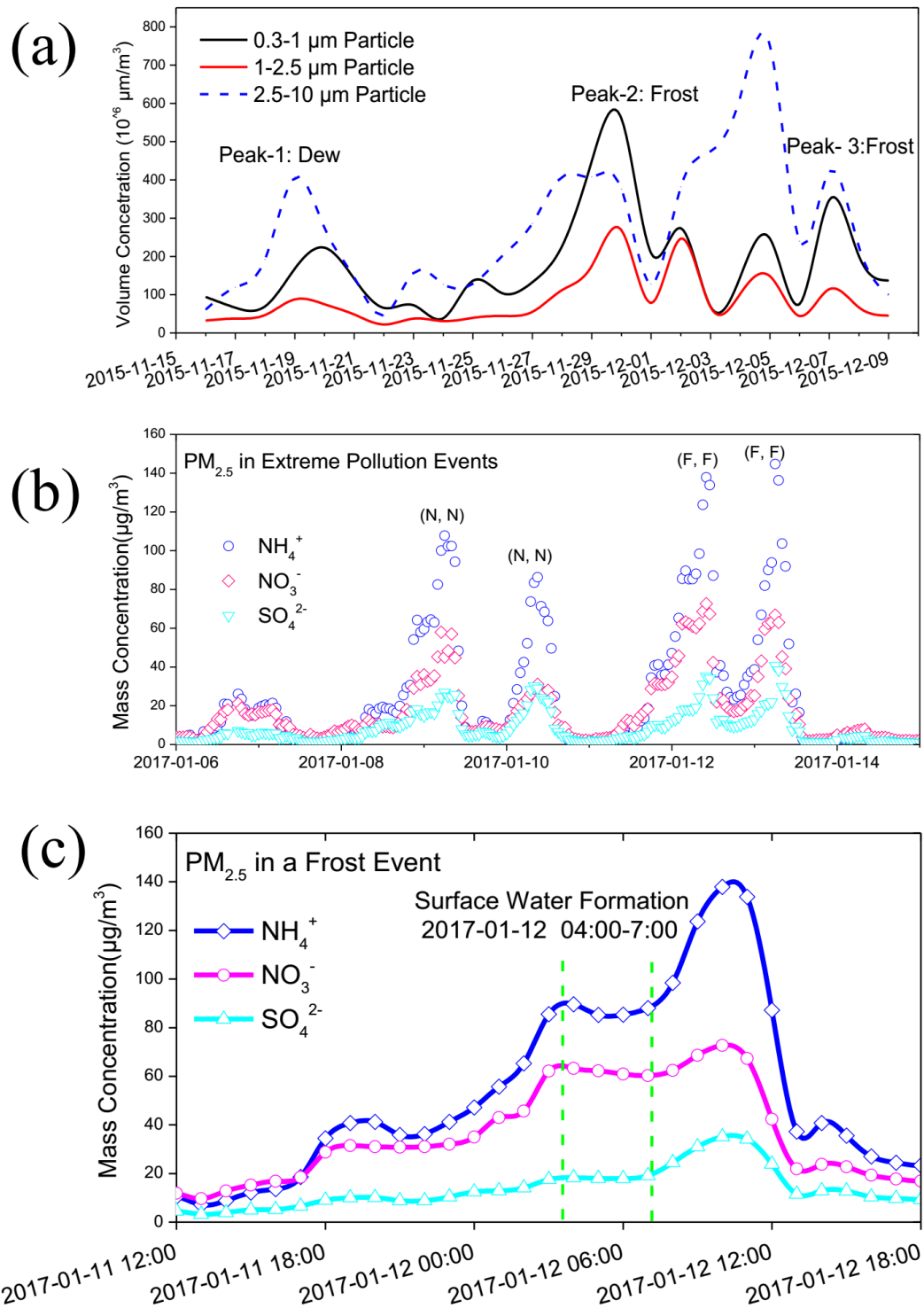

Fig. 9 Temporal variations during the periods of dew/frost events. (a) Temporal trends of volume concentration of particles, measured by OPS (TSI Corporation); (b)(c) Temporal trends of components in $PM_{2.5}$ measured by AIM-IC (Thermo Fisher Scientific).

### 3.2.2 SWF as a cleaner

The steps governing dew/frost composition are (1) deposition of solid particles, (2) dissolution of the soluble portion of the deposited particles by surface water, and (3) sorption of gases into the surface water[26]. However, there is no uniform and standard protocol for collecting and analyzing dew/frost solution. Researchers typically use an artificial film such as a polytetrafluoroethylene (PTFE) sheet and aluminum foil to collect surface water[27]. Nevertheless, these methods are based on artificial surfaces with different thermal properties from those of the natural surfaces. The dew volume per unit surface area was usually over-estimated. Therefore, the surface water collected by these films cannot represent the real compositions in dew/frost on leaves, soil, etc. To determine the real concentration of surface water on the natural surface, the surface itself has to be used as a collector[28]. An "ideal" condenser should thus be "grass-like", i.e. a light sheet thermally isolated from massive parts and from the ground, in an open area to radiate the energy and cooling[16]. Thus we regard the thin shrub leaf, covered by hydrophobic fatty substance and at 1-2 m height from the ground, as a perfect natural collector of condensed surface water.

Results have showed that surface water is a solution consists of dissolved gas and particles from atmospheric input, such as $SO_2$, $HNO_3$, $CO_2$, $Ca^{2+}$, $SO_4^{2-}$, and $NH_3$ absorbed by the surface water is re-emitted to the atmosphere[29]. The ionic concentrations are in general lower in dew than in rain[30]. Former study collected the surface water without pretreatment or cleaning of the collecting surface, since dry deposition of dust is inevitable in addition to the gas absorption[31]. Under natural conditions, it is impractical to find an absolutely clean leaf exactly exists during SWF. Here we washed out the surface water from the shrub leaves to analyze its inorganic composition without pretreatment, using $NH_4^+$, $NO_3^-$, $SO_4^{2+}$ as representative components of air pollutant input. The wash-out samples include Upside (Up) samples and Whole (Wh) samples, which represents washing upside leaf and washing the whole leaf, respectively. In addition to the dew and frost, we also wash the snow on the shrub canopy within one hours after the precipitation begun. Thus we get five groups of samples: (D-Up), (D-Wh), (F-Up), (F-Wh), (Snow) from wet leaves in the morning and dry leaves in the afternoon. When it comes to homogeneity test, only the data with the lowest $NH_4^+$ concentrations washed from the wet leaf did not meet the criteria of normal distribution. It means the different degree of $NH_3$ evaporation in the surface water due to different collecting time in the morning could cause a heterogeneity.

Assuming that the all of the $Ca^{2+}$ remain on the leaf after surface water evaporate, here we show the group mean values of normalized ions concentration in dew/frost samples for a concise view. See the Fig. 10(a)(b). The pH of surface water samples ranged from 5.86 to 8.93, indicating the mineralization in surface water by cations such as $Ca^{2+}$. Note that (D-Up) solutions has a much higher $NH_4^+$ and $SO_4^{2+}$ than other groups' samples. In addition to deposition of fine particles, liquid or droplet on the upside leaf absorbs more $SO_2$ and $NH_3$ gases from the overlying air than that in ice on the upside or in liquid on the reverse side of leaf. We can define Residual Percentage (RP) of each ions as concentration on dry leaf divided by concentration on wet leaf. See the Fig. 10(c). For $SO_4^{2+}$, RP-(D, Wh) and RP-(F, Wh) are 92% and 103%, respectively, suggesting that sulfate from the atmosphere remain on the leaf after surface water evaporation. However, for $NH_4^+$ and $NO_3^-$, RPs are 5-8% and 8-20%, respectively. We point out that >80% $NH_4NO_3$ is re-emitted to the atmosphere, while nearly all the sulfate remains on the leaf after surface water disappeared. Judging from the excessive $Ca^{2+}$ compared with sulfate and nitrate, we can see that the sulfate mainly exists as $CaSO_4$:

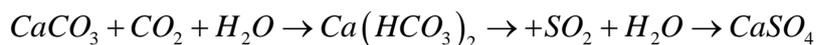

$$CaCO_3 + CO_2 + H_2O \rightarrow Ca(HCO_3)_2 \rightarrow +SO_2 + H_2O \rightarrow CaSO_4$$

To distinguish the aerosols input in surface water that deposited during and precede or after SWF events, we washed the dry leaves in a (N, N) event in 2017-01-05 after the SWF events occurred on the former days. When we standardize the ion concentration by $[Ca^{2+}]$, nitrate and sulfate in (N, N) samples account for 53%-70% of that in (D/F, D/F) samples, for an average of 62%. Aerosols deposited from air to canopy during SWF account for at least 40% of the nitrate and sulfate in the dew/frost solution. From the Fig. 10(b), assuming that [$Ca^{2+}$] in the surface water is 1 mmol/m$^2$ (meq), we estimate the total deposition flux of [$SO_4^{2-}$+$NO_3^-$] on the dry leaves as 0.8±0.1 mmol/m$^2$ (meq/$Ca^{2+}$). Therefore, the removal efficiency of thermophoresis deposition and surface water absorption for [$SO_4^{2-}$+$NO_3^-$] is estimated by ~0.3 mmol (per [$Ca^{2+}$] meq)/m$^2$(per leaf area). The $Ca^{2+}$ concentration in wash-out samples fluctuated from 0.08 to 0.25 mmol/m$^2$ due to the different choice of shrub leaves.

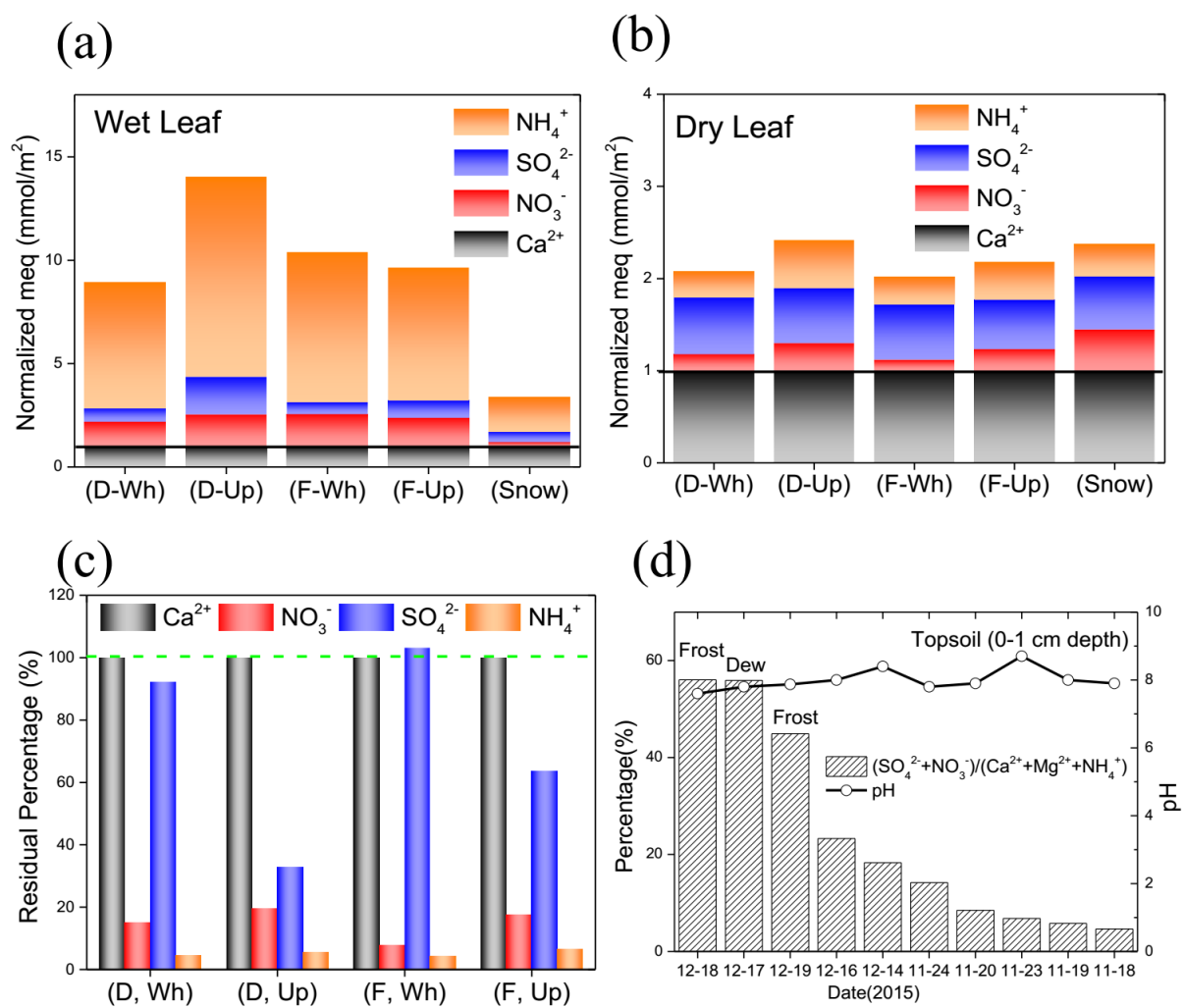

Fig. 10 Chemical compositions of wash-out samples from leaves: (a) Wash-out samples of dew/frost from wet leaf in the morning; (b) Wash-out samples from dry leaves in the afternoon in (D, D) and (F,

F) events; (c) Residual Percentage (%) = (Dry Leaf)/(Wet Leaf); (d) The ratio of ($SO_4^{2-}$+$NO_3^-$)/($Ca^{2+}$+$Mg^{2+}$+$NH_4^+$) in topsoil samples.

The surface water could reach up to 20 cm depth of the bare soil[32]. Airborne particle dissolved in the surface water could also permeate the topsoil. In the Fig. 10(d), we show the meq ratio of ($SO_4^{2+}$+$NO_3^-$)/($NH_4^+$ + $Ca^{2+}$+$Mg^{2+}$) in the 0-1 cm depth topsoil. We can see that in the (D, D) and (F, F) events, the ratio was more than 50%, while almost lower than 20% in the (N, N) events. With SWF on the ground, topsoil contains 3-5 times acid ions more than that without any surface water. However, the pH values of soil samples were >7 both in (D, D), (F, F) and (N, N) events, suggest that the input of acid components from the atmosphere couldn't reach the maximum neutralization capacity of the ground soil. Alkaline loess soil, which covers vast areas of N.E. China, contains $Ca^{2+}$ in large quantities, and calcium carbonate ($CaCO_3$) reacts with atmospheric $SO_2$, to form calcium sulphate ($CaSO_4$). Thus, even bare soil without vegetation may be a significant sink of acid components from atmosphere, acting as a neutralizing buffer for acidifying atmospheric deposition[8].

### 3.2.3 Brief summary: pollutant pool

Surface water formation is not only an indicator, but also a weak cleaner of the air pollution. During the winter with heating supply, anthropogenic sources emit a mass of water vapor and trace gases to the atmosphere, while enhanced moisture in the low temperature is favorable for the gas-particle formation of sulfate and nitrate. The ground soil is a pool of condensed surface water and deposited air pollutant. Considering that the alkaline loess soil in the northern China, the sand-dust aerosol could neutralize the acid component in the atmosphere.

# 4. Implication and Application

Challenge: (a) The aridity in the land, where the availability of surface water is limited, is substantially amplified by land-atmosphere feedbacks[33]. In the context of global warming, the evapotranspiration of the natural surface is enhanced. Satellite observations revealed that terrestrial evaporation had increased faster than precipitation in northern latitudes[34]. Water vapor content over land does not increase fast enough relative to the rapid warming there[35]. Therefore, the ground loses more moisture while the overlying air becomes drier[4,35,36]. (b) Human's urbanization processes have changed the land use type profoundly, with impervious surfaces replacing the vegetation and bare soil[37,38]. Urban land-cover is not conducive to the super-saturated condensation of surface water on that, which cause the soil moisture could not be replenished, thus aggravates the aridity. (c) People in northern China suffer from the severe condition of lacking fresh water resources, coupled with intensive air, soil and water pollution. The surface water is in poor quality, and cannot be treated as a reliable residential water.

Prospect: (a) It is ironic that the water vapor content is increasing while the atmosphere becomes drier due to global warming. As long as the substrate is hydrophilic and cooling enough, theoretically, the super-saturation is easier to reach and water vapor condensation could yield clean drinking water from the atmosphere, even in the desert area (see the pioneer study by Fathieh et al)[39]. Under the constant interfacial temperature difference, the super-saturation ratio is related to the coefficient of kinematic viscosity. To make this coefficient as large as possible, the joint efforts of multiple subjects such as engineering, material science, atmospheric science, and fluid mechanics must catch up the urgent demand. (b) It is also imperative to develop equipment for the utilization of latent heat[40-42]. The temperature difference generates electricity, known as the *Seeback effect*[43]. A material placed in a temperature gradient develops an electrical voltage between the hot and cold ends[43]. In the case of diurnal temperature range of 10 ℃, researchers have developed device is capable of producing electricity power of 350 mV and 1.3 mW from approximately 10 ℃ diurnal temperature differences [44]. A more reliable and environmental friendly facility is an effective remediation for heavy air pollution.

**Acknowledgements**

We would like to thank the support from the National Key Research and Development Program in China (No. 2016YFC0200504) and the Natural Science Foundation of China (Grant No. 41376087).

**Author Contributions**

Limin Feng wrote this manuscript. Yang Yu and Huan Xie participated in the experiments. Xiaohong Yao designed the experiments. Yujiao Zhu, Huiwang Gao and Xiaohong Yao modified this article.

**Competing Financial Interests**

The authors declare no competing financial interests.